\documentclass[conference]{IEEEtran}
\IEEEoverridecommandlockouts

\usepackage{cite}
\usepackage{amsmath,amssymb,amsfonts}
\usepackage{algorithmic}
\usepackage{graphicx}
\usepackage{textcomp}
\usepackage{xcolor}
\usepackage{tabularx}
\usepackage{booktabs}
\def\BibTeX{{\rm B\kern-.05em{\sc i\kern-.025em b}\kern-.08em
    T\kern-.1667em\lower.7ex\hbox{E}\kern-.125emX}}
\begin{document}

\title{Unmasking Shortcut Learning in IoT Intrusion Detection: A Forensic, Multi-Paradigm Evaluation of Feature Dependence and Data Leakage}
\author{\IEEEauthorblockN{Uday Shankar Roy}
\IEEEauthorblockA{\textit{Computer Science and Engineering} \\
\textit{Khulna University of Engineering \& Technology}\\
Khulna, Bangladesh \\
udayroycse@gmail.com}
\and
\IEEEauthorblockN{Mahbuba Jahan Minu}
\IEEEauthorblockA{\textit{Computer Science and Engineering} \\
\textit{Khulna University of Engineering \& Technology}\\
Khulna, Bangladesh \\
mjm1807120@gmail.com}
}
\maketitle

\begin{abstract}
Machine learning-based Network Intrusion Detection Systems often report near-perfect performance on IoT benchmarks. However, whether these models learn generalizable attack behavior or exploit spurious dataset shortcuts---such as static testbed IP/MAC addresses and chronological recording artifacts---remains an important question. We evaluate the CyberFlowIoT-GICAP benchmark, containing 3,617,388 flow records across 126 PCAP sessions with 849,395 benign flows. Four learning paradigms are evaluated across four feature configurations using PCAP-disjoint splits; LightGBM is additionally evaluated using conventional random-flow splitting. When only statistical flow behavior is used ($F_{\text{behav}}$), LightGBM (92.58\% $\pm$ 8.18\%), Random Forest (92.59\% $\pm$ 8.18\%), and Deep MLP (92.55\% $\pm$ 8.18\%) achieve nearly identical Macro-F1, indicating that performance is constrained by feature representation rather than model complexity. With raw timestamps ($F_{\text{tstamp}}$), tree-based models reach 99.28\% Macro-F1, while the linear model remains at 90.62\%, showing that nonlinear models can exploit dataset-specific temporal structure. Attack detectability is highly asymmetric: high-rate and active attacks maintain \textgreater 99.8\% recall from flow behavior alone in nonlinear models, whereas the DNS Beaconing drops from 27.78\% to 0.00\% recall when contextual features are removed. Conventional random-flow splitting increases attack recall by up to 14.00\%, highlighting the effect of placing flows from the same sessions in both training and test sets. We conclude with a 4-point protocol checklist for realistic IoT NIDS evaluation.
\end{abstract}


\section{Introduction}
The rapid expansion of Internet of Things (IoT) ecosystems across smart cities, industrial automation, and connected healthcare has expanded the cybersecurity threat landscape. Due to constrained computational resources and legacy firmware, IoT devices can be vulnerable to distributed attacks, including denial-of-service (DoS) floods, automated reconnaissance, man-in-the-middle (MITM) spoofing, and command-and-control (C2) beaconing. To protect these networks, machine learning (ML) and deep learning (DL) methods are widely used in Network Intrusion Detection Systems (NIDS) for automated, high-throughput classification of network flow telemetry.

Recent literature frequently reports near-perfect binary and multi-class classification performance on benchmark datasets. However, performance measured under controlled benchmark conditions may not translate directly to operational environments. This discrepancy can arise from \textit{shortcut learning}~\cite{geirhos2020shortcut,arp2022dos}, where statistical learning algorithms exploit predictive correlations that are not necessarily related to the underlying attack behavior. Such correlations may perform well under conventional i.i.d. evaluation but generalize poorly when the network environment, hosts, or recording conditions change.

In tabular network flow datasets, shortcut artifacts can arise through several structural mechanisms: (1)~\textbf{Host Identity Dependence}, where static IP/MAC addresses and OUIs associated with testbed devices provide predictive information unrelated to intrinsic traffic behavior; (2)~\textbf{Chronological Recording Artifacts}, where attacks recorded in separate time intervals allow models to associate calendar periods with attack labels; and (3)~\textbf{Intra-Session Data Leakage}, where conventional random flow splitting places flows from the same capture sessions in both training and test partitions, allowing session-specific characteristics to be shared across the evaluation boundary.

To systematically examine these mechanisms and characterize the generalization limits of ML-NIDS, this paper addresses three research questions:
\begin{itemize}
    \item \textbf{RQ1 (Feature Shortcuts \& Representation Bounds):} How do identity-inclusive ($F_{\text{all}}$), identity-scrubbed ($F_{-\text{ident}}$), and temporal ($F_{\text{tstamp}}$) configurations affect detection relative to pure behavior ($F_{\text{behav}}$), is performance bounded by model complexity or feature representation, and how do distinct paradigms exploit temporal artifacts?

    \item \textbf{RQ2 (Asymmetric Attack Detectability):} Which attack categories maintain invariant flow-level signatures, which become difficult to detect when identity and contextual features are removed, and how does model architecture affect their detectability?

    \item \textbf{RQ3 (Data Leakage \& Evaluation Protocols):} How does conventional random flow splitting affect performance compared with PCAP-disjoint evaluation, and what role does intra-session data sharing play in this difference?
\end{itemize}

\noindent\textbf{Key Findings and Contributions:} We conduct a systematic empirical investigation across the complete CyberFlowIoT-GICAP benchmark~\cite{martinez2026labelled}, comprising 3.61M flows across 126 PCAP sessions and dedicated benign traffic, using four learning paradigms (LightGBM, Random Forest, Deep MLP, and Linear SGD). Our main findings are:
(1)~\textbf{Empirical Representation Convergence:} On pure behavioral features ($F_{\text{behav}}$), LightGBM ($92.58\%$), Random Forest ($92.59\%$), and Deep MLP ($92.55\%$) achieve nearly identical Macro-F1, indicating that the evaluated performance ceiling is strongly constrained by the available feature representation rather than model complexity;
(2)~\textbf{Temporal Inductive Bias:} With raw timestamps ($F_{\text{tstamp}}$), tree-based models achieve $99.28\%$ Macro-F1, while the linear model remains at $90.62\%$, showing that nonlinear models can exploit dataset-specific temporal structure;
(3)~\textbf{Feature Dependence and Attack Asymmetry:} We identify a $99.50\%$ single-IP concentration among non-MITM attack flows and observe that DNS Beaconing recall falls to $0.00\%$ under pure behavioral features, while several high-rate and active attacks remain highly detectable;
and (4)~\textbf{Leakage Quantification and Evaluation Protocol:} We measure up to $+14.00\%$ attack-recall inflation under random flow splitting and derive a four-point protocol checklist for more realistic IoT NIDS evaluation.
\section{Related Work}
Evaluating machine learning systems in adversarial network environments is hindered by the semantic gap between statistical anomaly detection and operational realities, marked by high false-positive costs and non-stationary background traffic~\cite{sommer2010outside}. Meta-analyses by Arp et al.~\cite{arp2022dos} and Apruzzese et al.~\cite{apruzzese2023role} formalized pervasive methodological pitfalls in security ML, identifying \textit{sampling bias}, \textit{data snooping}, and \textit{spurious correlations} as leading drivers of non-replicable research. In tabular intrusion datasets---including KDD-CUP99~\cite{tavallaee2009detailed}, UNSW-NB15~\cite{moustafa2015unsw}, CICIDS2017~\cite{sharafaldin2018toward}, and TON\_IoT~\cite{alhawawreh2021toniot}---empirical studies by Sarhan et al.~\cite{sarhan2023towards} demonstrated that decision tree models routinely overfit to ephemeral metadata (e.g., transport ports, subnets) rather than learning intrinsic flow dynamics.

Beyond feature selection, forensic audits by Engelen et al.~\cite{engelen2021troubleshooting} and Liu et al.~\cite{liu2022error} documented systematic construction errors and labeling artifacts in enterprise benchmarks, including labeling logic historically dependent on attacker IP addresses and recording intervals. Our forensic audit (Section~III) reveals similar IP- and timestamp-dependent artifacts in a modern IoT benchmark, demonstrating that such vulnerabilities persist well beyond legacy datasets.

In the IoT/IIoT domain, Reddy and Kumar~\cite{reddy2026datasetcentric} surveyed 91 studies and concluded qualitatively that near-perfect IDS accuracies stem from class imbalance and evaluation bias rather than genuine detection capability. However, their contribution is a qualitative taxonomy; it lacks empirical feature ablations, multi-paradigm model comparisons, and forensic ground-truth auditing.

In the broader ML literature, Geirhos et al.~\cite{geirhos2020shortcut} formalized the theoretical foundations of shortcut learning, showing that statistical optimizers can exploit simple predictive signals even when they lack causal relevance. Taken together, prior work has either (i) audited legacy enterprise datasets without connecting errors to a multi-paradigm learning study~\cite{engelen2021troubleshooting,liu2022error}, (ii) surveyed IoT dataset realism qualitatively without controlled ablation~\cite{reddy2026datasetcentric}, or (iii) established shortcut theory without network telemetry grounding~\cite{geirhos2020shortcut}. A controlled, multi-paradigm empirical study combining forensic auditing, systematic feature ablation, and session-level PCAP holdout on a modern IoT benchmark has remained unexplored.
\section{Dataset Audit}

\subsection{Benchmark Composition}
We evaluate the complete CyberFlowIoT-GICAP benchmark~\cite{martinez2026labelled}, which comprises 3,617,388 bidirectional flow records spanning 849,395 dedicated background benign flows and 2,767,993 flows collected across 126 distinct PCAP capture sessions. As detailed in Table~\ref{tab:pcap_inventory}, the benchmark contains seven attack categories exhibiting severe class imbalance and heterogeneous capture compositions.

\begin{table}[htbp]
\vspace{-8pt}
\caption{Complete Forensic Inventory across 126 PCAP Captures}
\label{tab:pcap_inventory}
\centering
\footnotesize
\setlength{\tabcolsep}{3.2pt}
\begin{tabular}{lrrrr}
\toprule
\textbf{Category} & \textbf{PCAPs}$^*$ & \textbf{Total Flows} & \textbf{Attacks} & \textbf{Atk \%} \\
\midrule
BENIGN (Dedicated) & --- & 849,395 & 0 & 0.00\% \\
Denial of Service (DoS) & 43 (35/8) & 960,657 & 593,576 & 61.79\% \\
ARP Spoofing (MITM) & 8 (8/0) & 70,915 & 61,042 & 86.08\% \\
Nmap Reconnaissance & 30 (30/0) & 130,229 & 51,155 & 39.28\% \\
SQL Injection (SQLi) & 6 (6/0) & 105,107 & 33,814 & 32.17\% \\
DNS Beaconing & 22 (13/9) & 1,072,347 & 1,611 & 0.15\% \\
MQTT Injection & 11 (10/1) & 396,914 & 486 & 0.12\% \\
FuerzaBruta & 6 (6/0) & 31,824 & 223 & 0.70\% \\
\midrule
\textbf{Total / Overall} & \textbf{126 (108/18)} & \textbf{3,617,388} & \textbf{741,907} & \textbf{20.51\%} \\
\bottomrule
\multicolumn{5}{@{}p{\linewidth}@{}}{\vspace{2pt}\footnotesize $^*$Parentheses denote (Mixed / Pure) PCAP counts. Mixed = captures with both benign and attack flows; Pure = fully benign captures.}
\end{tabular}
\vspace{-10pt}
\end{table}

\subsection{IP Address Distribution}
Analyzing the source IP addresses of all attack flows reveals two distinct structural patterns in the benchmark. In the six non-MITM attack categories (totaling 680,865 attack flows), 99.50\% of flows (677,429 flows) originated from a single machine (\texttt{192.168.8.120}). When IP addresses are included in training, this concentration creates an opportunity for models to exploit source identity as a shortcut rather than learning intrinsic flow behavior. In contrast, ARP Spoofing (MITM) involved distributed traffic across multiple endpoints: the IoT Data Hub (\texttt{192.168.8.121}), a primary intercepted communication endpoint, accounted for 36.96\% of attack flows (22,564 flows), while 30 poisoned victim devices (\texttt{192.168.8.150} through \texttt{192.168.8.179}) produced 62.77\% of attack flows (38,316 flows) due to traffic interception and forwarding by the attacker node (\texttt{192.168.8.120}).

\subsection{Timestamp and Recording Artifacts}
Examining the flow start timestamps reveals that all attack experiments were recorded one after another in separate, non-overlapping time windows (dedicated benign traffic was recorded first, followed on subsequent days by DoS, MITM, MQTT, Nmap, Brute Force, SQL Injection, and DNS Beaconing). Because each attack category was recorded during an isolated calendar interval, raw timestamps provide a strong shortcut for separating traffic along the temporal axis rather than detecting genuine intrusion dynamics.

\section{Methodology}
\begin{figure}[ht]
\vspace{-8pt}
\centering
\includegraphics[width=\columnwidth, height=3.5in, keepaspectratio]{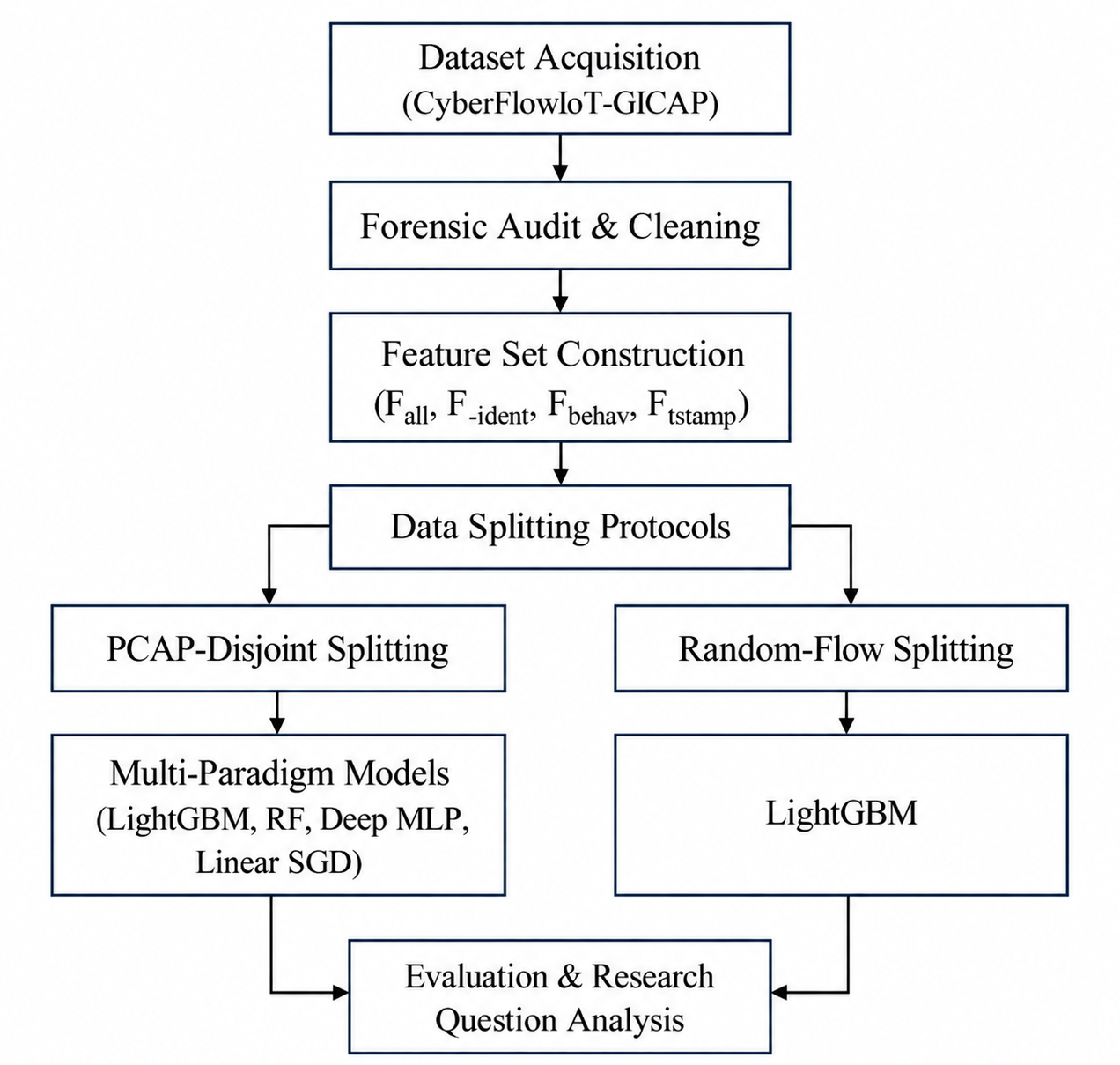}
\caption{Overall experimental workflow}
\label{fig:workflow}
\vspace{-8pt}
\end{figure}

The overall experimental workflow is summarized in Fig.~\ref{fig:workflow}, spanning dataset acquisition and forensic auditing through feature construction, dual data splitting protocols, and multi-paradigm evaluation.
\subsection{Feature Taxonomy}
The raw dataset contains 98 attributes per flow record. We permanently exclude 17 administrative metadata and ground-truth boundary annotations (such as row IDs, PCAP hashes, and \texttt{gt\_start\_ms}) to prevent label leakage, leaving 81 valid network telemetry features. 

To systematically evaluate shortcut learning, we organize these features into five constituent groups: six host identifiers spanning Layer-2/Layer-3 addresses (\texttt{src\_ip}, \texttt{dst\_ip}, \texttt{src\_mac}, \texttt{dst\_mac}, \texttt{src\_oui}, \texttt{dst\_oui}); two transport ports (\texttt{src\_port}, \texttt{dst\_port}); 12 application and protocol metadata fields (e.g., \texttt{protocol}, \texttt{user\_agent}, \texttt{client\_fingerprint}); 55 behavioral flow dynamics capturing directional traffic volumes, flow duration, packet sizes, inter-arrival times, and cumulative TCP flags; and six absolute epoch timestamps (e.g., directional \texttt{first\_seen\_ms}, \texttt{last\_seen\_ms}). As summarized in Table~\ref{tab:feature_taxonomy}, these groups form our four evaluation configurations: the identity-inclusive 75-feature baseline ($F_{\text{all}}$), an identity-scrubbed configuration ($F_{-\text{ident}}$, 69 features), pure behavioral dynamics ($F_{\text{behav}}$, 55 features), and temporal augmentation ($F_{\text{tstamp}}$, 61 features combining behavior with timestamps).
\begin{table}[htbp]
\vspace{-8pt}
\caption{Landmark Feature Taxonomy}
\label{tab:feature_taxonomy}
\centering
\footnotesize
\setlength{\tabcolsep}{3.5pt}
\begin{tabularx}{\columnwidth}{lcX}
\toprule
\textbf{Feature Set} & \textbf{Count} & \textbf{Included Feature Groups} \\
\midrule
$F_{\text{all}}$ & 75 & 55 Behavioral, 12 App/Proto, 2 Ports, 6 Host IDs \\
$F_{-\text{ident}}$ & 69 & 55 Behavioral, 12 App/Proto, 2 Ports \\
$F_{\text{behav}}$ & 55 & 55 Behavioral Features (Volumes, PIATs, TCP Flags, etc.) \\
$F_{\text{tstamp}}$ & 61 & 55 Behavioral, 6 Absolute Epoch Timestamps \\
\bottomrule
\end{tabularx}
\vspace{-6pt}
\end{table}

\subsection{Multi-Paradigm Models}
A critical question is whether different learning paradigms rely on the same shortcut features (such as static IP addresses and timestamps) or if certain architectures are naturally resistant. To investigate this, we select four diverse model families representing major paradigms in machine learning (summarized in Table~\ref{tab:model_taxonomy}):

\noindent\textbf{LightGBM (Gradient Boosting)~\cite{ke2017lightgbm}:} Boosted decision trees are the standard benchmark for tabular network traffic. We include LightGBM to test how well it learns attack patterns, and whether its step-by-step tree building makes it prone to exploiting IP and timestamp shortcuts.

\noindent\textbf{Random Forest (Bagged Trees)~\cite{breiman2001random}:} Instead of building trees sequentially, Random Forest trains independent trees on random subsets of data and averages their predictions. We include it to test if averaging multiple trees helps the model avoid memorizing specific dataset shortcuts.

\noindent\textbf{Deep MLP (Neural Networks)~\cite{paszke2019pytorch}:} To evaluate continuous, gradient-based learning alongside decision trees, we implement a multi-layer perceptron trained with AdamW optimizer. This allows us to test whether neural networks generalize better to unseen network captures or fall for the same dataset shortcuts.

\noindent\textbf{Linear SGD (Convex Linear Baseline)~\cite{pedregosa2011scikit}:} We include a basic linear classifier as a baseline. This allows us to see if complex models are truly necessary for flow detection, and to test how a simple linear boundary responds to timestamp artifacts.
\begin{table}[htbp]
\vspace{-8pt}
\caption{Evaluated Model Architectures \& Hyperparameters}
\label{tab:model_taxonomy}
\centering
\footnotesize
\setlength{\tabcolsep}{3.5pt}
\begin{tabularx}{\columnwidth}{llX}
\toprule
\textbf{Model} & \textbf{Paradigm} & \textbf{Core Hyperparameters} \\
\midrule
LightGBM & GBDT & $N=100$, $\text{lr}=0.05$, $\text{num\_leaves}=31$, $\text{jobs}=10$ \\
Random Forest & Bagging & $N=100$, $\text{max\_samples}=0.25$, $\text{depth}=20$ \\
Deep MLP & Deep NN & $128\text{-}64\text{-}32\text{-}2$, BatchNorm, Dropout 0.2, AdamW \\
Linear SGD & Linear & $\text{log\_loss}$, L2, $\alpha=10^{-4}$, $\text{max\_iter}=1000$ \\
\bottomrule
\end{tabularx}
\vspace{-10pt}
\end{table}

\subsection{Data Splitting Protocols}
To assess generalization beyond individual capture sessions and quantify data leakage, we evaluate two complementary splitting protocols:

\noindent\textbf{PCAP-Disjoint Split (Primary Protocol):} For each attack category, PCAPs are partitioned 80\% train / 20\% test across three independent seeds (42, 123, 2026), ensuring zero capture overlap ($\mathcal{P}_{\text{train}} \cap \mathcal{P}_{\text{test}} = \emptyset$). PCAPs are assigned randomly without preferentially selecting mixed or pure captures. Dedicated background benign traffic (849,395 flows, \texttt{source\_pcap}=\texttt{NULL}) is independently split 80/20. This zero-leakage protocol governs our 48 primary benchmark trials across four learning paradigms and four feature configurations.

\noindent\textbf{Stratified Random-Flow Split (Conventional Baseline):} As a conventional benchmark, flows are randomly partitioned into an 80/20 stratified split ($N_{\text{train}}=2,893,911$; $N_{\text{test}}=723,477$), disregarding PCAP boundaries and allowing flows from the same capture session to appear in both partitions. This protocol is evaluated using LightGBM across all four feature configurations and three seeds (12 audit trials) to isolate the performance inflation caused by intra-session leakage.

\subsection{Evaluation Metrics}
Due to severe class imbalance (79.49\% benign traffic, with attacks like MQTT and DNS Beaconing representing under 0.15\% of flows), raw accuracy is deceptive. We adopt Macro-F1 (the unweighted average of benign and attack F1-scores) as our primary metric to treat both classes equally. Additionally, we evaluate Per-Attack Recall across all seven attack categories to uncover specific detection blind spots. All experiments report the exact $\text{Mean} \pm \text{Standard Deviation}$ across three random seeds.
\subsection{Implementation Details}
The experimental pipeline was implemented in Python. The benchmark evaluates four model architectures (LightGBM, Random Forest, Deep MLP, and Linear SGD) across four feature configurations ($F_{\text{all}}$, $F_{-\text{ident}}$, $F_{\text{behav}}$, and $F_{\text{tstamp}}$). The evaluation comprises 60 distinct experimental trials executed across three fixed random seeds (42, 123, and 2026). Under the primary PCAP-disjoint splitting protocol, all four architectures are trained and evaluated across all four feature sets over the three seeds ($4 \times 4 \times 3 = 48$ trials). For comparison against conventional flow splitting, LightGBM is additionally evaluated across all four feature sets under stratified random flow-level splitting using the same three seeds ($1 \times 4 \times 3 = 12$ trials). All performance metrics are reported as the mean $\pm$ standard deviation across the three runs.

\section{Results and Discussion}

\subsection{RQ1: Feature Shortcuts \& Temporal Inductive Bias}
Table~\ref{tab:master_results} reports the macro-F1 scores across all four architectures and four feature sets under PCAP-disjoint holdout.

\begin{table}[htbp]
\caption{Master Multi-Paradigm Macro-F1 Benchmark Across Feature Sets (Mean $\pm$ SD \% Across 3 Seeds)}
\label{tab:master_results}
\centering
\footnotesize
\setlength{\tabcolsep}{3.2pt}
\begin{tabular}{lllll}
\toprule
\textbf{Model} & \multicolumn{1}{c}{$\mathbf{F_{\text{all}}}$} & \multicolumn{1}{c}{$\mathbf{F_{-\text{ident}}}$} & \multicolumn{1}{c}{$\mathbf{F_{\text{behav}}}$} & \multicolumn{1}{c}{$\mathbf{F_{\text{tstamp}}}$} \\
\midrule
LightGBM   & $88.36{\pm}6.42$ & $88.07{\pm}6.50$ & $92.58{\pm}8.18$ & $99.28{\pm}0.42$ \\
RF         & $89.11{\pm}6.49$ & $89.09{\pm}6.48$ & $92.59{\pm}8.18$ & $97.37{\pm}1.17$ \\
Deep MLP   & $88.66{\pm}6.42$ & $88.65{\pm}6.43$ & $92.55{\pm}8.18$ & $84.07{\pm}13.03$ \\
Linear SGD & $89.86{\pm}6.42$ & $89.33{\pm}6.45$ & $90.62{\pm}7.00$ & $90.62{\pm}7.00$ \\
\bottomrule
\end{tabular}
\vspace{-12pt}
\end{table}

\noindent\textbf{Behavioral Features Outperform Identity-Loaded Baselines:}
Across all four models, restricting features to pure statistical flow behavior ($F_{\text{behav}}$) consistently improves generalization over the 75-feature baseline $F_{\text{all}}$ (e.g., LightGBM: $92.58\%$ vs. $88.36\%$). Notably, removing host identifiers alone ($F_{-\text{ident}}$) yields negligible change ($88.07\%$ vs. $88.36\%$). The performance surge occurs only when transitioning to $F_{\text{behav}}$, which strips transport ports (\texttt{src\_port}, \texttt{dst\_port}) and application-layer metadata. This shows that port numbers and protocol fields, not just IP addresses, cause models to overfit to the testbed. 

Furthermore, on pure behavior ($F_{\text{behav}}$), LightGBM ($92.58\% \pm 8.18\%$), Random Forest ($92.59\% \pm 8.18\%$), and Deep MLP ($92.55\% \pm 8.18\%$) converge to identical Macro-F1 scores within $0.04\%$. This empirical parity indicates that tree boosting, bagging, and deep neural networks are equally constrained by the feature space: when shortcuts are removed, the performance ceiling is bounded by the representation itself, rather than by architectural complexity.

\noindent\textbf{Contrasting Inductive Biases on Timestamps (Trees vs. Linear Models):}
Augmenting behavioral features with raw epoch timestamps ($F_{\text{tstamp}}$) causes LightGBM and Random Forest to surge to $99.28\%$ and $97.37\%$ Macro-F1, respectively. Because decision trees construct axis-aligned decision boundaries, they readily conjoin temporal boundaries with behavioral thresholds:
\begin{equation}
f_{\text{tree}}(\mathbf{x}, t) = \sum_{k=1}^K w_k \cdot \mathbb{I}\left(t_{\text{start}}^{(k)} \le t \le t_{\text{end}}^{(k)}\right) \prod_{j \in \mathcal{S}_k} \mathbb{I}\left(x_j \le \theta_{j}^{(k)}\right)
\end{equation}
This allows tree ensembles to use timestamps as localized temporal gates. By first isolating the discrete calendar windows during which specific attack campaigns were recorded, the model immediately eliminates background benign traffic recorded outside those intervals (including the 849,395 dedicated benign flows), and then uses localized behavioral splits to separate the remaining interleaved benign and attack flows within that window.

In sharp contrast, Linear SGD achieves identical performance under $F_{\text{behav}}$ and $F_{\text{tstamp}}$ ($90.62\% \pm 7.00\%$). A convex linear model evaluates a single global hyperplane:
\begin{equation}
f_{\text{linear}}(\mathbf{x}, t) = \mathbf{w}^T \mathbf{x} + w_t t + b
\end{equation}
Because attack campaigns occurred in scattered, non-contiguous calendar intervals, a single linear direction cannot partition disjoint temporal epochs. Empirical inspection of the trained SGD coefficients reveals that timestamp features receive modest but non-zero weights (mean $|w_t| \approx 0.26$, with $|w_t| \le 0.76$), which are substantially smaller than the top behavioral features (up to $|w| \approx 5.92$). However, because a linear hyperplane lacks the axis-aligned gating of tree splits, these weights provide no measurable net improvement to Macro-F1 ($90.62\%$ across both configurations), in sharp contrast to the massive surge seen in tree ensembles.

Finally, unlike the tree ensembles, Deep MLP's Macro-F1 degrades from $92.55\% \pm 8.18\%$ under $F_{\text{behav}}$ to $84.07\% \pm 13.03\%$ under $F_{\text{tstamp}}$. This decline is driven by a collapse in DoS recall from $99.89\% \pm 0.04\%$ (under $F_{\text{behav}}$, Table~\ref{tab:per_attack_fbehav}) to $72.52\% \pm 47.57\%$ (under $F_{\text{tstamp}}$, Table~\ref{tab:shortcut_drilldown}).

\subsection{RQ2: Asymmetric Attack Vulnerability \& DNS Collapse}
Table~\ref{tab:per_attack_fbehav} presents cross-model recall across individual attack categories on pure behavioral features ($F_{\text{behav}}$).

\begin{table}[htbp]
\vspace{-8pt}
\caption{Cross-Model Per-Attack Recall on Pure Behavior ($F_{\text{behav}}$, Mean $\pm$ SD \% Across 3 Seeds)}
\label{tab:per_attack_fbehav}
\centering
\footnotesize
\setlength{\tabcolsep}{2.0pt}
\begin{tabular}{lllll}
\toprule
\textbf{Attack} & \multicolumn{1}{c}{\textbf{LightGBM}} & \multicolumn{1}{c}{\textbf{RF}} & \multicolumn{1}{c}{\textbf{MLP}} & \multicolumn{1}{c}{\textbf{SGD}} \\
\midrule
DoS          & $99.94{\pm}0.05$ & $99.96{\pm}0.02$ & $99.89{\pm}0.04$ & $94.01{\pm}10.05$ \\
Nmap Recon   & $99.92{\pm}0.09$ & $99.96{\pm}0.06$ & $99.83{\pm}0.07$ & $91.91{\pm}6.98$ \\
SQLi         & $99.97{\pm}0.01$ & $99.99{\pm}0.01$ & $99.97{\pm}0.01$ & $99.77{\pm}0.14$ \\
FuerzaBruta  & $92.75{\pm}8.76$ & $98.41{\pm}2.75$ & $66.03{\pm}13.64$ & $28.82{\pm}24.66$ \\
MQTT         & $88.56{\pm}7.95$ & $88.84{\pm}7.64$ & $86.47{\pm}8.30$ & $45.90{\pm}4.33$ \\
ARP Spoofing & $35.66{\pm}0.09$ & $35.66{\pm}0.08$ & $35.52{\pm}0.14$ & $35.48{\pm}0.11$ \\
DNS Beacon   & $0.00{\pm}0.00$  & $0.00{\pm}0.00$  & $0.00{\pm}0.00$  & $0.00{\pm}0.00$ \\
\bottomrule
\end{tabular}
\end{table}

\noindent\textbf{Invariant Signatures in High-Rate and Active Attacks:}
Active and high-volume attack categories achieve near-perfect recall across all non-linear models on pure behavior ($F_{\text{behav}}$): DoS ($>99.8\%$), Nmap Recon ($>99.8\%$), and SQL Injection ($>99.9\%$). Even the convex linear classifier maintains over $91\%$ recall on these categories. High-rate flooding (DoS), rapid connection sweeps (Nmap), and characteristic payload/TCP sequences (SQL Injection) each produce genuine, invariant behavioral signatures that models learn reliably without relying on testbed metadata shortcuts.

\noindent\textbf{MITM Detection Ceiling and Architecture-Dependent Shortcut Exploitation:}
For ARP Spoofing MITM, all four learning architectures hit an identical detection ceiling of approximately $35.6\%$ on pure behavior ($F_{\text{behav}}$). Because flow statistics only aggregate packet volumes and timings, man-in-the-middle forwarding flows blend closely with standard local network communication.
\begin{table}[htbp]
\vspace{-6pt}
\caption{Cross-Model Shortcut Sensitivity on Critical Attacks (Mean $\pm$ SD \%)}
\label{tab:shortcut_drilldown}
\centering
\footnotesize
\setlength{\tabcolsep}{2.5pt}
\begin{tabular}{lccc}
\toprule
\textbf{Model} & \textbf{DoS} ($F_{\text{tstamp}}$) & \textbf{ARP Spoof} ($F_{\text{tstamp}}$) & \textbf{DNS Beacon} ($F_{\text{all}}$) \\
\midrule
LightGBM    & $99.96{\pm}0.02$  & $99.50{\pm}0.37$   & $27.78{\pm}4.81$ \\
RF          & $99.96{\pm}0.02$  & $47.59{\pm}12.67$  & $27.78{\pm}4.81$ \\
Deep MLP    & $72.52{\pm}47.57$ & $65.72{\pm}11.67$  & $27.78{\pm}4.81$ \\
Linear SGD  & $94.01{\pm}10.06$ & $35.48{\pm}0.11$   & $0.00{\pm}0.00$  \\
\bottomrule
\end{tabular}
\vspace{-6pt}
\end{table}

Importantly, testing timestamp augmentation ($F_{\text{tstamp}}$, Table~\ref{tab:shortcut_drilldown}) reveals that shortcut exploitation on ARP Spoofing MITM is highly architecture-dependent. LightGBM leverages timestamps to surge to $99.50\%$ recall, and Deep MLP reaches $65.72\%$. In contrast, Random Forest only reaches $47.59\%$, while Linear SGD remains completely unchanged at $35.48\%$. This demonstrates that the degree to which a model exploits chronological recording windows depends directly on its underlying inductive bias and capacity to partition non-monotonic calendar time.

\noindent\textbf{Collapse of Covert DNS Beaconing:}
Under the 75-feature baseline ($F_{\text{all}}$), LightGBM, Random Forest, and Deep MLP all achieve identical DNS Beaconing recall of $27.78\% \pm 4.81\%$ by memorizing the dedicated attacker IP address (\texttt{192.168.8.120}). However, once identity shortcuts are eliminated ($F_{\text{behav}}$), recall collapses to exactly $0.00\% \pm 0.00\%$ across all four learning paradigms, while Linear SGD detects $0.00\%$ across all feature sets.

This collapse stems from extreme traffic dilution. Forensic auditing reveals that the 1,611 covert DNS beacon flows are diluted across 1,070,736 benign background flows ($0.15\%$ attack density). Each beacon flow consists of standard UDP port 53 packets (60 to 80 bytes) with standard query durations. Without static IP addresses, isolated single-flow statistics cannot distinguish covert periodic queries from legitimate background DNS lookups. Reliable detection of covert IoT threats requires stateful multi-flow aggregation across time windows rather than isolated static flow classification.
\subsection{RQ3: Evaluation Bounds \& Intra-Session Data Leakage}
Table~\ref{tab:splitting_audit} compares PCAP-disjoint holdout against conventional stratified random flow splitting on LightGBM across all four feature configurations.

\begin{table}[htbp]
\vspace{-8pt}
\caption{Evaluation Protocol Benchmark on LightGBM: PCAP-Disjoint vs. Random Flow Splitting (\% Across 3 Seeds)}
\label{tab:splitting_audit}
\centering
\footnotesize
\setlength{\tabcolsep}{3.0pt}
\begin{tabular}{lcccccc}
\toprule
& \multicolumn{3}{c}{\textbf{Macro-F1 (\%)}} & \multicolumn{3}{c}{\textbf{Attack Recall (\%)}} \\
\cmidrule(lr){2-4} \cmidrule(lr){5-7}
\textbf{Feature Set} & \textbf{PCAP} & \textbf{Rand} & $\mathbf{\Delta}$ & \textbf{PCAP} & \textbf{Rand} & $\mathbf{\Delta}$ \\
\midrule
$F_{\text{all}}$     & 88.36 & 87.99 & $-0.37$ & 80.60 & 93.49 & $\mathbf{+12.89}$ \\
$F_{-\text{ident}}$  & 88.07 & 88.00 & $-0.07$ & 79.65 & 93.65 & $\mathbf{+14.00}$ \\
$F_{\text{behav}}$   & 92.58 & 88.65 & $-3.93$ & 89.25 & 94.45 & $+5.19$ \\
$F_{\text{tstamp}}$  & 99.28 & 99.57 & $+0.29$ & 99.86 & 99.42 & $-0.45$ \\
\bottomrule
\end{tabular}
\vspace{-2pt}
\end{table}

Conventional random flow splitting inflates attack recall by up to $+14.00\%$ on $F_{-\text{ident}}$ and $+12.89\%$ on $F_{\text{all}}$ (Table~\ref{tab:splitting_audit}). Because 108 of the 126 attack PCAPs contain interleaved benign traffic, random splitting scatters flows from the same capture session into both training and test sets. This allows models to memorize intra-session flow distributions rather than generalizable attack dynamics.

However, this recall gain comes at the cost of elevated false alarms: on $F_{\text{behav}}$, attack precision drops from $86.78\%$ under PCAP holdout to $73.44\%$ under random splitting, pulling down overall Macro-F1 ($-3.93\%$). Finally, under $F_{\text{tstamp}}$, the leakage delta is negligible ($\Delta\text{Rec} = -0.45\%$), because calendar timestamps provide a global shortcut that operates independently of session boundaries.

\subsection{Practical Implications and Evaluation Guidelines}
Our results indicate that volumetric and active attacks are detectable from behavior alone, low-rate covert threats collapse once identity and metadata shortcuts are removed, and conventional random splitting substantially inflates measured performance through intra-session leakage. Addressing the covert-attack gap likely requires moving beyond isolated flow classification toward stateful, multi-flow temporal aggregation (e.g., session-level query-entropy or host-interaction graph modeling); we did not implement or evaluate such approaches here, and mark this as a direction for future work.

Two limitations bound these findings: non-MITM attacks originated overwhelmingly from a single testbed IP ($99.50\%$), so spatial generalization across multi-subnet deployments remains untested; and covert categories comprised as little as $0.12\%$--$0.15\%$ of their capture files, too few instances to train or evaluate stateful sequence models within this study.

To reduce shortcut learning and non-replicable benchmark claims in future NIDS research, we formalize a 4-point protocol checklist:
\begin{enumerate}\setlength{\itemsep}{1pt}\setlength{\parskip}{0pt}
    \item \textbf{Mandatory Capture-Level Holdout:} Benchmarks must partition data at the physical capture file or device level ($\mathcal{P}_{\text{train}} \cap \mathcal{P}_{\text{test}} = \emptyset$) to prevent intra-session data leakage.
    \item \textbf{Removal of Host Identifiers and Timestamps:} Feature sets should exclude static IPs, MACs, OUIs, and absolute epoch timestamps prior to training to reduce models' ability to substitute testbed-specific artifacts for genuine behavioral learning.
    \item \textbf{Multi-Seed and Multi-Paradigm Validation:} Evaluations should report exact Mean and Standard Deviation across multiple random seeds, evaluating both tree ensembles and neural models to expose inductive bias vulnerabilities.
    \item \textbf{Mandatory Per-Attack Recall Reporting:} Research should report granular per-attack recall matrices alongside aggregate Macro-F1 to detect asymmetric vulnerabilities and covert attack collapse.
\end{enumerate}

\section{Conclusion}
This study presents an exhaustive empirical investigation into shortcut learning and multi-paradigm generalization in IoT intrusion detection, evaluating four learning paradigms across four feature configurations. Our results show that LightGBM, Random Forest, and Deep MLP converge within $0.04$ percentage points on pure flow behavior ($92.55\%$--$92.59\%$ Macro-F1), indicating that detection capacity is bounded primarily by the feature representation rather than model complexity. Tree ensembles exploit disjoint calendar recording windows to reach up to $99.28\%$ Macro-F1 when raw timestamps are available, whereas the linear classifier derives no comparable benefit ($90.62\%$, unchanged from pure behavior). Covert DNS Beaconing collapses to $0.00\%$ recall across all four paradigms once identity shortcuts are removed, and conventional random flow splitting inflates attack recall by up to $+14.00\%$ relative to realistic PCAP-disjoint holdout. These findings motivate the four-point evaluation protocol proposed in Section~V: capture-level holdout, removal of host identifiers and timestamps, multi-seed and multi-paradigm validation, and mandatory per-attack recall reporting. Future NIDS research should focus on session-level analysis to detect low-rate covert threats that remain difficult to identify from individual network flows alone.


\bibliographystyle{IEEEtran}
\bibliography{references}

\end{document}